\documentstyle[psfig]{mn0}

\def\simg{\mathrel{\rlap{\raise 0.511ex \hbox{$>$}}{\lower 0.511ex \hbox{$\sim$}}}}
\def\siml{\mathrel{\rlap{\raise 0.511ex \hbox{$<$}}{\lower 0.511ex \hbox{$\sim$}}}}
\def\ks{\,{\rm ks}} \def\day{\,{\rm d}} \def\eV{\,{\rm eV}} \def\keV{\,{\rm keV}}
\def\GHz{\,{\rm GHz}} \def\Meszaros{M\'esz\'aros~}

\begin{document}

\parskip 3pt
\topmargin 5pt

\title[GRB afterglow 130427A]
    {An external-shock model for GRB afterglow 130427A}

\author[Panaitescu, Vestrand \& Wo\'zniak]{A. Panaitescu, W.T. Vestrand, P. Wo\'zniak \\
       Space \& Remote Sensing, MS B244, Los Alamos National Laboratory,
       Los Alamos, NM 87545, USA}

\maketitle

\begin{abstract}
%\begin{small}
\\
 The complex multiwavelength emission of GRB afterglow 130427A (monitored in the radio up to 10 days,
 in the optical and X-ray until 50 days, and at GeV energies until 1 day) can be accounted for by a 
 hybrid reverse-forward shock synchrotron model, with inverse-Compton emerging only above a few GeV.
 The high ratio of the early optical to late radio flux requires that the ambient medium is a wind
 and that the forward-shock synchrotron spectrum peaks in the optical at about 10 ks. The latter has 
 two consequences: the wind must be very tenuous and the optical emission before 10 ks must arise from 
 the reverse-shock, as suggested also by the bright optical flash that Raptor has monitored during the
 prompt emission phase ($<$ 100 s). The VLA radio emission is from the reverse-shock, the Swift X-ray 
 emission is mostly from the forward-shock, but the both shocks give comparable contributions to the 
 Fermi GeV emission. The weak wind implies a large blast-wave radius ($8\, t_{day}^{1/2}$ pc),
 which requires a very tenuous circumstellar medium, suggesting that the massive stellar progenitor 
 of GRB 130427A resided in a super-bubble. 
% The afterglow apparent size ($20\, t_{day}^{3/4}\, \mu$as) and radio flux expected from the forward-shock 
% (around 0.06 mJy at 15 GHz, constant until 200 day) may allow this source to be resolved with radio interferometry. 
%\end{small}
\end{abstract}


\section{Introduction}

 Gamma-Ray Burst (GRB) 130427A may well be the burst with the most comprehensive afterglow follow-up, its
multiwavelength monitoring covering radio, optical, X-ray, and $\gamma$-ray frequencies, and extending
from seconds to tens of days after trigger. The X-ray {\sl prompt} emission (up to 100 s) was accompanied by
the second brightest optical flash, monitored by Raptor (Vestrand et al 2013), with the optical afterglow 
light-curve displaying a steepening at 300 s and a flattening at 10 ks. The Swift X-ray light-curve 
(X-ray light-curve repository -- Evans et al 2009) is consistent with a single power-law from 500 s to 5 Ms. 
The Fermi-LAT $\gamma$-ray light-curve (Tam et al 2013) displays a peak at 10--20 s, simultaneous 
with the optical flash peak, and a steepening at 550--800 s (Zhu et al 2013). The VLA radio light-curves 
(Laskar et al 2013) display a slow decay at 1-10 day.

 With such a rich dataset, GRB afterglow 130427A demands a theoretical interpretation, done here in the
framework of the external-shock model (\Meszaros \& Rees 1997) where some relativistic ejecta,
produced by the black-hole resulting from the core-collapse of a massive star, drive a {\sl forward-shock}
into the ambient medium while the ejecta are energized by the {\sl reverse-shock}. The synchrotron and
inverse-Compton emissions from both shocks are calculated assuming that electrons and magnetic field
acquire a certain fraction of the post-shock energy. The shock-accelerated electrons are assumed to have
a power-law distribution with energy (hence the synchrotron and inverse-Compton spectra are also power-laws),
with a break at the cooling energy (where the radiative-loss timescale equals the shock age). 
 
 Analytical treatments for the forward-shock emission have been provided by \Meszaros \& Rees (1997),
Sari, Piran \& Narayan (1998), Waxman, Kulkarni \& Frail (1998), Granot, Piran \& Sari (1999), Wijers \& 
Galama (1999), Chevalier \& Li (2000), Panaitescu \& Kumar (2000), and for the reverse-shock by Kobayashi 
(2000). Both shocks have been studied with 1-dimensional hydrodynamical codes by Panaitescu \& \Meszaros 
(1998) and Kobayashi \& Sari (2000), the former focusing on the two-shock synchrotron and inverse-Compton 
emission, the latter on the dynamics of the shocks.

 To model the multiwavelength emission of GRB afterglow 130427A, we employ a 1-dimensional code that follows 
the ejecta--medium interaction, with the dynamics of each shock calculated from conservation of energy and 
using the shock jump-conditions (Blandford \& McKee 1976). After the onset of deceleration, the dynamics of 
the forward-shock is determined by the ejecta initial energy, injected energy (Rees \& \Meszaros 1998), 
and ambient medium density. The dynamics of the reverse-shock is determined by that of the shocked fluid and 
two properties of the incoming ejecta: their energy and Lorentz factor. Here, we consider that the ejecta add
energy to the blast-wave as a power-law in observer time and that they have a single Lorentz factor.  
The self-absorption and cooling frequencies of the synchrotron spectrum and the inverse-Compton parameter 
are calculated self-consistently from the electron distribution and the magnetic field strength (Panaitescu 
\& \Meszaros 2000). Radiative losses are also calculated from the electron distribution, but they are
negligible for the following best-fit models.
The emissions from both shocks are integrated over their motion and over the angle at which the fluid moves 
relative to the direction toward the observer. More details about this numerical model and its application
to the multiwavelength emission of ten GRB afterglows are given in Panaitescu (2005). 

\vspace*{-3mm}
\section{Reverse-forward (external) shock model}

\subsection{Closure relations for forward-shock light-curves suggest a homogeneous medium}

 The choice of the model features that may accommodate the temporal decay of the broadband emission 
of GRB afterglow 130427A starts with the X-ray light-curve because its temporal decay index ($F \propto
t^{-\alpha}$) and spectral slope ($F_\nu \propto t^{-\beta}$) are the best determined: $\alpha_x = 1.36 \pm 0.01$ 
at 20 ks -- 5 Ms (Fig \ref{homog}) and $\beta_x = 0.79 \pm 0.16$ at mean time 24 ks (Kennea et al 2013). 
These lead to $\alpha_x - 1.5\beta_x = 0.18 \pm 0.24$, which is compatible with the value expected (zero) 
for the synchrotron emission from the forward-shock interacting with a homogeneous medium and for the X-ray
being below the cooling frequency $\nu_c$ of the synchrotron spectrum. 

 As the optical flux also decays at that time, the optical must be above the peak energy $\nu_p$ of the 
synchrotron spectrum, hence optical and X-ray are in the same spectral regime: $\nu_p < \nu_o (2 \eV) < 
\nu_x (10 \keV) < \nu_c$. Consequently, the intrinsic afterglow optical flux can be calculated from the X-ray flux: 
$F_o = F_x (\nu_o/\nu_x)^{\beta_x}$. For instance, the observed $F_{10\,keV} (54 \ks) = 1.4\, \mu$Jy implies 
that $F_{2 eV} (54 \ks) = 1.2$ mJy, which is a factor 2.5 larger than the measured $F_{2 eV} (53 \ks) = 0.47$ mJy, 
requiring $A_R = 1$ mag of dust extinction in the host galaxy. 

 The 10 keV -- 100 MeV spectral slope $\beta_{xg} (43 \ks) \simeq 0.89 \pm 0.09 \simg \beta_x$ indicates that 
$\nu_c$ is well above 10 keV. That the LAT flux decays slower than in the X-ray ($\alpha_g = 1.22 \pm 0.09$ 
at 500 s -- 50 ks) indicates that $\nu_c$ is below the LAT range (otherwise, for $\nu_c > 100$ MeV,
the model expectation is $\alpha_x = \alpha_g$) and that the electron radiative cooling is dominated by 
inverse-Compton scatterings (otherwise, for synchrotron-dominated electron cooling, $\nu_c \propto t^{-1/2}$ and
$\alpha_g - \alpha_x = - 0.5 (d\log \nu_c/d\log t) = 1/4$, incompatible with the observed $\alpha_g \siml \alpha_x$). 
More exactly, for a Compton-dominated electron cooling, the decay index of the synchrotron flux above $\nu_c$ 
is $\alpha = 3p/4 - 1/(4-p) = 1.27$, which matches well the observed $\alpha_g$, with $p = (4\alpha_x + 3)/3 = 2.8$
being the exponent of the power-law distribution of electrons with energy ($dN/d\epsilon \propto \epsilon^{-p}$) 
that is required by the forward-shock model, given the measured flux decay index $\alpha_x$ below the cooling frequency.

 In summary, the optical, X-ray, and $\gamma$-ray fluxes of GRB afterglow 130427A, their decay indices, and the 
X-ray spectral slope, require that $\nu_p < \nu_o < \nu_x < \nu_c < \nu_g$, if the afterglow emission is synchrotron
from the forward-shock.

\subsection{External medium is not homogeneous}

 Under the assumption that the two microphysical parameters of the forward-shock ($\epsilon_B$ and $\epsilon_i$) 
that quantify the post-shock fractional energy in the magnetic field and in electrons are constant, the 
forward-shock synchrotron light-curve at any frequency below the optical can be easily 
calculated from the optical light-curve, using the expected evolution of the synchrotron peak flux ($F_p = const$) 
and peak energy ($\nu_p \propto t^{-3/2}$) for a homogeneous medium. If $\nu_p$ crosses the optical at 
some time $t_o$, yielding an optical flux $F_o$, then the radio flux at frequency $\nu_r < \nu_p$ is 
\begin{equation}
 F_r (t) = F_p (\nu_r/\nu_p)^{1/3} = F_o (t_o) (t/t_o)^{1/2} (\nu_r/\nu_o)^{1/3} \propto t^{1/2}
\label{radio}
\end{equation}
Here, $F_o(t_o) \simeq 5\, (t_o/10 \ks)^{-\alpha_o}$ mJy is the intrinsic optical light-curve after 10 ks 
(corrected for the above-inferred host extinction of $A_R = 1$ mag) and $\alpha_o = 1.36$ 
(the forward-shock model requires that $\alpha_o = \alpha_x$). 
The largest $t_o$ required by equation (\ref{radio}) arises from the radio measurement with the highest
$\nu_r^{1/3} t^{1/2}/F_r$; taking the $F_{36 \GHz} (9.7 \day) = 0.43$ mJy measurement as an upper
limit for the forward-shock radio flux, implies $t_o \simg 23$ ks. 

 This means that, for the forward-shock emission (that accommodates the observed optical flux) not to 
exceed the measured radio fluxes, the synchrotron peak should cross the optical at 23 ks. 
Conversely, if the synchrotron peak crossed the optical before 23 ks, then the synchrotron flux
from the forward-shock would violate VLA measurements. That may be avoided if the magnetic field parameter 
$\epsilon_B$ decreases (roughly as $t^{-1}$), and if energy injection in the forward-shock is allowed 
(to match the optical and X-ray flux decays at 1--10 day, which are faster when $\epsilon_B$ decreases), 
but this scenario requires fine-tuning and we do not pursue it.

%\subsection{Numerical proof against a homogeneous medium}
% A possible shortcoming of the forward-shock interacting with a homogeneous medium model is the break
%displayed by the LAT light-curve at $\sim$ 300 s, which cannot be explained by this model. As discussed 
%above, the slower decay of the LAT flux (than of the XRT flux) requires that the synchrotron cooling
%frequency is between 10 keV and 100 MeV and that electron cooling is inverse-Compton--dominated, so
%that $\nu_c$ increases in time. Thus there is no spectral break whose crossing of the LAT range could
%yield a light-curve break. An origin of the LAT light-curve break in the dynamics of the forward-shock
%requires an increase in the shock deceleration rate, as for a jet-break (but the post jet-break flux decays
%are too slow) or for a shock entering a region where the medium density $n$ increases(decreases) with radius 
%faster(slower) than before (so that the change in the evolution of $\nu_c$ across that density discontinuity
%yields a light-curve break at $\nu > \nu_c$).
 
\begin{figure}
\centerline{\psfig{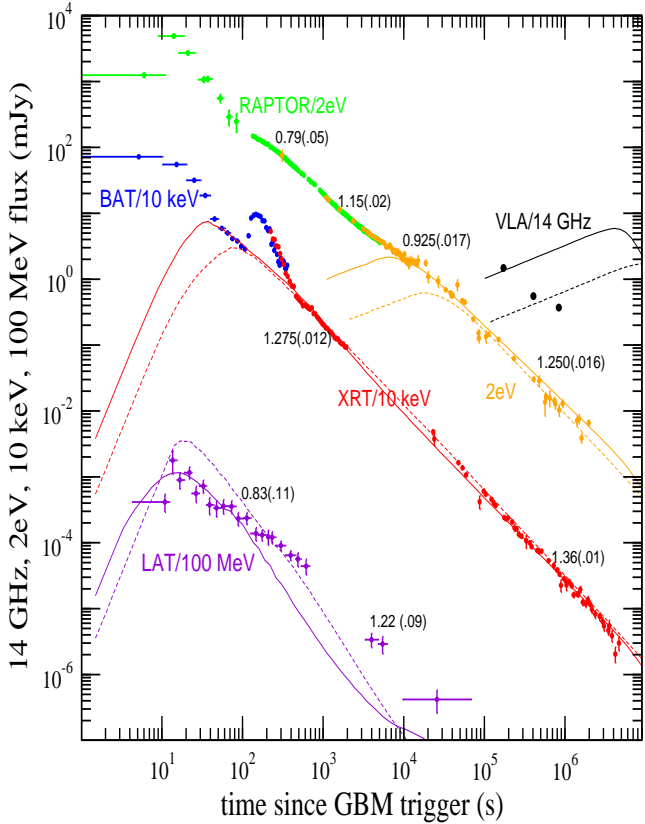}}
\caption{ Multiwavelength light-curves for GRB afterglow 130427A and the {\sl synchrotron 
   forward-shock/homogeneous-medium} model best-fit to the radio, optical after 10 ks, X-ray after 50 s, 
   and $\gamma$-ray measurements (the second X-ray pulse at 100--400 s is not included in the fit). 
   Numbers adjacent to light-curves give the local power-law flux decay index (and its $1\sigma$ uncertainty). \newline
   {\sl Solid} lines are for a model excluding radio measurements; the peak of the synchrotron spectrum crosses 
   the R-band (2 eV) at 10 ks and the model {\sl over-predicts} some of the VLA radio measurements 
   at 1--10 day and 1--90 GHz (as expected from eq \ref{radio}). 
   The parameters of this model are: ejecta initial kinetic energy $E_0 = 3.10^{53}$ erg/sr, ejecta initial 
   Lorentz factor $\Gamma_0 = 850$ (to yield a 20 s peak for the 100 MeV flux, when the ejecta deceleration begins), 
   ambient medium density $n = 2.10^{-3} {\rm cm^{-3}}$, 
%   jet opening $\theta > 0.15$ rad (so that a jet-break is not seen until a few Ms), 
   magnetic-field parameter $\epsilon_B = 10^{-4}$, electron minimum-energy parameter $\epsilon_i = 0.11$, 
   index of electron power-law distribution with energy $p = 2.5$, host dust-extinction $A_V = 1.3$. \newline
   {\sl Dashed} lines are for a model including the radio measurements, which forces the peak of the synchrotron 
   spectrum to be higher (crossing the optical at 20 ks). This model still over-predicts some radio measurements 
   as well as the $\gamma$-ray flux measured by LAT during the prompt phase. Its parameters are similar to the
   solid lines model, the most notable difference being $\epsilon_i = 0.23$.
  }
\label{homog}
\end{figure}

 Fig \ref{homog} illustrates the failure of the forward-shock synchrotron model with a homogeneous 
medium to accommodate simultaneously the radio, optical, X-ray, and $\gamma$-ray fluxes of GRB afterglow
130427A: while it can explain the optical afterglow emission after 10 ks and the X-ray flux at 50 s -- 5 Ms
(excluding the second GRB pulse, which is a feature that cannot be accounted for by any type of external
shock), this model over-predicts either the lowest or the highest frequency data. 

\subsection{Wind-like medium and forward-shock emission for the late optical afterglow} 
\label{sypeak}

 The closure relation $\alpha_x - 1.5\beta_x = 0.18 \pm 0.24$ is also compatible with the forward-shock model 
expectation for a wind-like medium (with an $n \propto r^{-2}$ particle density distribution with radius) 
and for X-ray below the cooling frequency, provided that there is an energy injection in the forward-shock 
that slows its deceleration and the decay of the afterglow X-ray flux. If that energy injection is parametrized 
as $E \propto t^e$ in observer time (a power-law flux decay requires that the dynamics of the forward-shock is 
a power-law in observer time), then $\alpha_x - 1.5\beta_x = [1+(\beta_x + 1)e]/2$, from where 
$e = 3 - 2(\alpha_x +1)/(\beta_x +1) = 0.36 \pm 0.23$. 

 Numerically, we find that the best-fit to the X-ray emission after 500 s (including all the GeV data and the 
optical after 10 ks) has $e = 0.30$ and that forward-shock energy should increase by a factor $E_i/E_0 = 3$ 
until $t_e \simeq 1$ Ms, to account for the observed X-ray flux decay. That means that the energy added to the 
forward-shock mitigates its deceleration after $t_i = t_e (E_0/E_i)^{1/e} = 20$ ks.

 It is important to note that the forward-shock interacting with a wind-like medium does not produce more radio 
emission than measured because the synchrotron peak flux decreases as $F_p \propto t^{-1/2}$ (instead of being 
constant, as for a homogeneous medium). The evolution of the synchrotron peak energy is the same as for a 
homogeneous medium ($\nu_p \propto t^{-3/2}$), hence the radio flux expected from the optical emission is 
\begin{equation}
 F_r (t) = F_p (\nu_r/\nu_p)^{1/3} = F_o (t_o) (\nu_r/\nu_o)^{1/3} \propto t^0
\label{radios2}
\end{equation}
Then, $F_o(t_o) \simeq 2 (t_o/10 \ks)^{-1.36}$ mJy and $F_{36 \GHz}(9.7 \day) = 0.43$ mJy require that the time 
when the synchrotron peak crosses the optical is $t_o \simg 3$ ks. That brief flattening seen in the optical 
light-curve at 10 ks could be due to $\nu_p$ crossing the optical and is compatible with $t_o \simg 3$ ks.

\begin{figure}
\centerline{\psfig{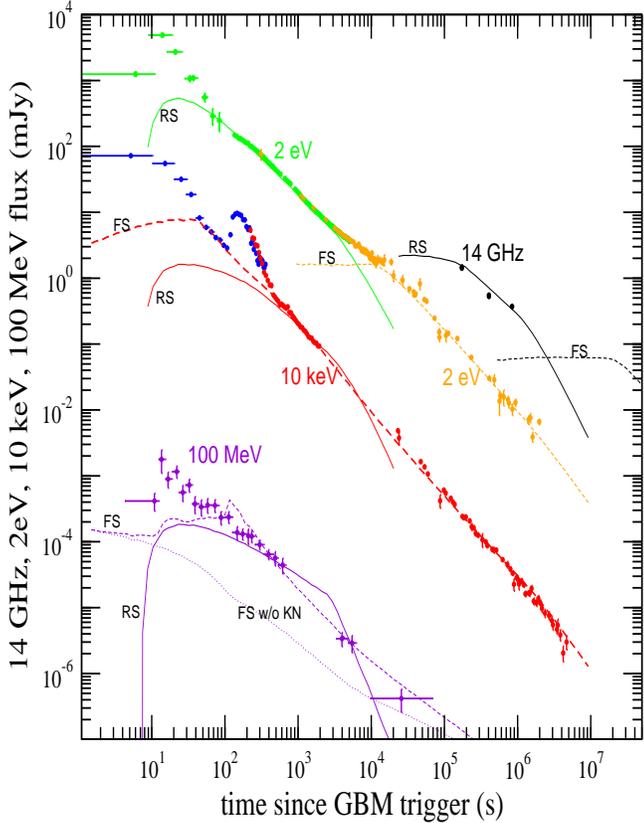}}
\caption{ Best-fit with a hybrid {\sl reverse(RS)/forward(FS)} shock model for the broadband emission of GRB 
  afterglow 130427A and for a {\sl wind-like} medium. 
  Solid lines are for the reverse-shock light-curves, dashed lines for the forward-shock; the dotted line shows 
  the 100 MeV forward-shock flux when the Klein-Nishina effect is ignored.  \newline
  {\bf RS} parameters:
  {\sl at 10s--3ks}: leading ejecta energy $E_0 = 10^{54}$ erg/sr, incoming ejecta energy $E_i^{(1)} = 4.10^{53}$ 
  erg/sr ($E_i^{(1)} < E_0$, hence the dynamics of the forward-shock is not affected by this first energy injection 
  episode), incoming ejecta Lorentz factor $\Gamma_i = 1800$, wind-density parameter $A = 0.004$ (see text for why 
  such a low density is required), $\epsilon_B = 10^{-3}$, $\epsilon_i = 0.011$, $p=2.0$;
% (RS-electrons injected until 10 ks cool adiabatically afterward and their emission decays fast)
  {\sl at 3ks--1Ms}: same $E_0$ and $A$ as above, $E_i^{(2)} = 4.10^{54}$ erg/sr ($E_i^{(2)} > E_0 + E_i^{(1)}$, 
  thus this second energy injection mitigates the blast-wave deceleration), $\Gamma_i = 3000$, $\epsilon_B = 
  2.10^{-5}$, $\epsilon_i = 0.016$, $p=2.3$ . 
  The RS optical, X-ray, and GeV emissions (not shown) for the latter injection episode are dimmer than from the FS
  at same time. \newline
%  {\sl Note}: the RS crossing the ejecta that yield the energy injection required by the optical and X-ray 
%  flux decays can account for the 1--10 day VLA radio emission, but its microphysical parameters are different 
%  than at 10--3000 s, when the RS produces the optical afterglow. 
  {\bf FS} parameters: same $E_0$, $E_i^{(2)}$ and $A$ as above, $\epsilon_B = 2.10^{-5}$, $\epsilon_i = 0.14$, $p=2.3$ .
  (Laskar et al 2013 have found almost the same low wind density, with similar $\epsilon_e$ and $p$, but an 
  $\epsilon_B$ close to equipartition and an ejecta kinetic energy that is 1000 times smaller).  
 }
\label{wind}
\end{figure}

 The best-fit to the optical data after 10 ks, the X-ray after 500 s, and all GeV measurements, with the
forward-shock emission and for a wind-like medium is shown in Fig \ref{wind}, with a sequence of spectra 
shown in Fig \ref{spek}.
The $\chi^2_\nu = 5.7$ for 135 dof of that best-fit makes it statistically unacceptable; the GeV fit has 
the largest $\chi^2_\nu = 7.1$ for 9 points, closely followed by the optical fit's $\chi^2_\nu = 6.3$ 
for 39 points, with the largest contribution to the fit's $\chi^2$ arising from the X-ray data,
$\Delta \chi^2 = 392$ for 79 points. The model light-curves follow well all flux trends and relative 
intensities except the brightness of the prompt emission until 50 s, but cannot describe well the early GeV
light-curve and cannot capture the fluctuations in the X-ray and optical measurements (after 10 ks,
optical data are from different instruments).

\subsection{A very tenuous wind}
\label{tenouswind}

 Compared to the parameters inferred for other afterglows by modelling their multiwavelength emission,
the wind density of the best-fit shown in Fig \ref{wind} is very small, but not unprecedented (Chevalier, 
Li \& Fransson 2004). Its parameter, $A = 0.003$, corresponds to a stellar mass-loss rate --to-- terminal 
wind-velocity ratio ($\dot{M}/v$) that is 300 smaller than for a typical Wolf-Rayet (WR) star (as the 
progenitor of long bursts with an associated Type Ic supernovae), for which $\dot{M} = 10^{-5} 
\dot{M}_{-5} M_\odot/$yr and $v = 10^8 v_8$ cm/s. 
The reason for that low density is the requirement that the synchrotron peak crosses the optical after 3 ks 
and matches the optical flux detected at that time. For $z=0.34$ and for the fluid moving directly toward the 
observer, the forward-shock synchrotron peak energy and peak flux are 
\begin{equation}
 h \nu_p (10 \ks) = 0.5\, E_{54}^{1/2} \epsilon_{B,-5}^{1/2} \epsilon_{i,-1}^2\, {\rm eV} 
\end{equation}
\begin{equation}
  F_p (10 \ks) = 240\, E_{54}^{1/2} \epsilon_{B,-5}^{1/2} A \; {\rm mJy} 
\end{equation}
Imposing that $\nu_p (10 \ks) = \nu_o = 2$ eV and $F_p (10 \ks) = F_o (10 \ks) = 2$ mJy, yields
\begin{equation}
  E_{54}^{1/2} \epsilon_{B,-5}^{1/2} \epsilon_{i,-1}^2  = 0.011  \; , \; 
  E_{54}^{1/2} \epsilon_{B,-5}^{1/2} A  = 2.6 \times 10^{-5}
\end{equation}
Taking the ratio of these two equations leads to $A = 2.3 \times 10^{-3} \epsilon_{i,-1}^2$. The $\epsilon_i$
parameter that quantifies the typical electron energy corresponds to a total electron energy that is a fraction
$\epsilon_e = (p-1)/(p-2) \epsilon_i$ of the post-shock energy. Equipartition with protons sets an upper limit,
$\epsilon_e \leq 1/2$, thus $\epsilon_i \leq 0.12$ for $p = 2.32$, from where $A \siml 0.003$. 

 This wind density is about 20 times lower than the lowest value measured (Nugis \& Lamers 2000) for Galactic 
WR stars and indicates a low mass loss-rate combined with a high wind velocity. Provided that can happen at 
the end of a WR's life, it has a strong consequence on the medium in which that star resides, as following. 
Owing to low wind density and high ejecta kinetic energy, the forward-shock that fits the late time broadband 
emission of GRB afterglow 130427A is highly relativistic, having $\Gamma \simeq 80 (t/1 \day)^{-1/4}$, hence 
the shock radius is $R_a = 2\Gamma^2 ct = 8 (t/1 \day)^{1/2}$ pc. Requiring that $R_a$ at the latest observation 
epoch (50 day) is less than the size of the bubble blown by a WR star during its $10^6$ yr lifetime, 
$R_s = 36\, (\dot{M}_{-5} v_8^2/n_0)^{1/5}$ pc (cf. Castor, McCray \& Weaver 1975), with $n$ the 
medium density around the star, we find that $n \siml 5.10^{-4} v_8^3 (t_s/50 \day)^{-5/2}\, {\rm cm^{-3}}$ 
for a wind with $\dot{M}_{-5}/v_8 = 0.004$, where $t_s$ is the observer-frame epoch when the afterglow shock
encounters the wind termination shock. 
Such a low ambient density suggests that the progenitor of GRB 130427A occurred in a supper-bubble 
(Scalo \& Wheeler 2001) blown by many preceding supernovae.

\subsection{Novel details of the forward-shock model, both related to the high-energy LAT emission}

\begin{figure*}
\centerline{\psfig{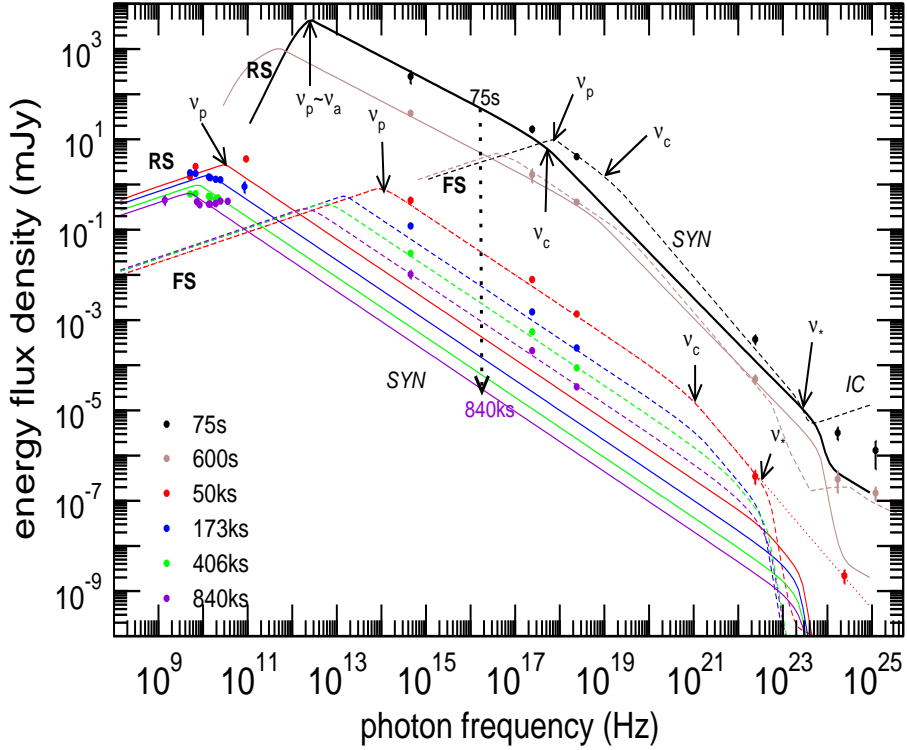}}
\caption{ Sequence of spectra for the reverse-forward shock model of Fig \ref{wind}, at the epochs indicated 
   in the legend. Data at same epoch and the corresponding model spectrum have the same colour, solid lines are 
   for the reverse-shock, dashed for the forward-shock. The spectral breaks indicated are: $\nu_a$ 
   (self-absorption frequency), $\nu_p$ (peak frequency, for electrons of typical post-shock energy, 
   parametrized by $\epsilon_i$), $\nu_c$ (cooling frequency, corresponding to electrons whose radiative 
   cooling timescale equals the shock's age), and $\nu_*$ (cut-off frequency, corresponding to electrons that 
   lose their energy during one gyration). 
   The dotted line shows the fit to the 50 ks $\gamma$-ray spectrum obtained with the synchrotron FS 
   emission if the electron acceleration timescale were much shorter than the gyration time. 
   The forward-shock inverse-Compton emission emerges above the synchrotron cut-off, yielding a harder
   spectrum above a few GeV and accounting for the higher-energy LAT emission until several ks.
 }
\label{spek}
\end{figure*}

 There are two interesting facts related to the LAT emission produced by the forward-shock synchrotron
model shown in Figs \ref{homog} -- \ref{spek}. First is that the scattering of the synchrotron
emission (at the peak of the spectrum) by the forward-shock electrons (of typical energy) occurs near 
the Klein-Nishina (KN) regime.
When the electron cooling is dominated by inverse-Compton scatterings (i.e. Compton parameter $Y>1$),
the cooling frequency satisfies $\nu_c \propto Y^{-2}$. Inclusion of the KN effect reduces the Compton 
$Y$ parameter, thus, taking into account the KN effect, increases $\nu_c$ and the synchrotron flux at 
$\nu > \nu_c$: $F_\nu \propto \nu_c^{1/2} \propto Y^{-1}$. In other words, the synchrotron emission 
from fast-cooling electrons increases when a competing radiative process (inverse-Compton) is reduced 
(by inclusion of the KN effect). 

 For the forward-shock best-fit to GRB afterglow 130427A, the LAT range is above $\nu_c$ and $Y>1$;
inclusion of the KN effect reduces $Y$ by about 10 and increases the 100 MeV 
flux by an order of magnitude (see Fig \ref{wind}). 
Furthermore, as the electrons at the peak of their distribution with energy enter and exit the KN regime,
the synchrotron light-curve at 100 MeV displays more structure than when the KN effect is ignored.

 The second is that radiative cooling during one gyration time limits the energy that electrons 
acquire through first-order Fermi acceleration to a corresponding synchrotron characteristic energy 
$h\nu_* \simeq 60 (z+1)^{-1}\, \Gamma/(Y+1)$ MeV, independent of the magnetic field $B$. 
For the best-fit parameters given in Fig \ref{wind}, the forward-shock has $\Gamma (1 \ks) = 240$ 
and $Y (1 \ks) 18$, so the maximal synchrotron energy is $h\nu_* (1 \ks) \simeq 600$ MeV 
(see synchrotron spectrum cut-off in Fig \ref{spek}). 
At earlier times, that cut-off is higher, but the inverse-Compton emission from the forward-shock 
takes over above 2 GeV (as shown by the $t=75$ s spectrum) and can account for the higher-energy LAT 
emission until about 10 ks, after which the inverse-Compton flux is too low. 

 Interestingly, a hardening of the LAT spectrum above several GeV was identified by Tam et al (2013), 
from $\beta_g^{(low)} = 1.2 \pm 0.1$ at 0.1--5 GeV to $\beta_g^{(high)} = 0.4 \pm 0.2$ at 5--100 GeV. 
Tam et al (2013) have proposed that the harder high-energy component is inverse-Compton, although we
find that the observed spectrum above 5 GeV is softer than the model expectation $\beta_g^{(high)} = -1/3$, 
corresponding to the GeV range being below the peak of the upscattered spectrum.

%Interestingly, if that synchrotron cut-off energy due to radiative losses during acceleration were not taken
%into account, then the synchrotron forward-shock emission would have a spectrum that matches the entire
%LAT range (100 MeV -- 100 GeV) spectrum (see Fig \ref{spek}), which suggests that electrons are accelerated
%on timescale shorter than for the Fermi mechanism, perhaps by the high electric fields generated in 
%plasma-instabilities (e.g. Weibel instability - Medvedev \& Loeb 1999). 

\subsection{Reverse-shock emission for the radio and early optical afterglow}

 The estimation of the expected radio emission given in equation (\ref{radios2}) led to the conclusion that 
the forward-shock cannot account for the optical afterglow emission prior to $\sim 10$ ks. Also, the flat radio 
light-curve arising from the forward-shock interacting with a wind cannot account for the radio emission at 1--10 day,
which is slowly decaying. Both these emissions are attributed to the reverse-shock (see also Laskar et al 2013),
as discussed below. We note that, after 10 ks, the existence of a reverse-shock is required by the energy 
injection into the forward-shock required by the measured decay index of the X-ray flux.

 The radio data are contemporaneous with the higher energy (optical, X-ray, and $\gamma$-ray) afterglow emission 
accommodated by the forward-shock, thus, for the calculation of the reverse-shock emission, the dynamical parameters 
$E_0$, $E_i^{(2)}$, $e$, and $A$ are fixed at the values determined from the forward-shock best-fit. The free parameters
of the reverse-shock are the Lorentz factor $\Gamma_i$ of the incoming ejecta (which sets the post-shock energy 
density) and the three microphysical parameters ($\epsilon_B$, $\epsilon_i$, and $p$) that determine the synchrotron 
spectrum. The best-fit obtained with the reverse-shock emission to the 1--10 day radio data is shown in Figs 
\ref{wind} and \ref{spek}. Unfortunately, it has a large $\chi^2_\nu = 25$ for 25 dof, because it underestimates 
the radio flux above 50 GHz. As shown in Fig \ref{spek}, those radio data cannot be explained by the forward-shock 
either, if its microphysical parameters are constant.
%Interestingly, $\epsilon_B$ and $p$ of the RS best-fit to the radio emission are the same as for the FS model 
%to the higher frequency emission, but $\epsilon_i$ is ten times smaller. 
Requiring the same microphysical parameters for both the reverse and forward shocks yields a much worse radio data
fit, with $\chi^2_\nu = 47$.

 The best-fit to the early optical emission with a reverse-shock includes also the earlier X-ray data and all 
GeV data, to ensure that the reverse-shock emission does not exceed what was observed. 
Again, the dynamical parameters $E_0$ and $A$ are fixed to the values obtained for the forward-shock, but
the energy $E_i^{(1)}$ carried by the incoming ejecta arriving at the blast-wave prior to 10 ks is only weakly 
constrained by the forward-shock fit to the optical and X-ray data after 10 ks, which sets an upper 
limit $E_i^{(1)} < E_0$. With free micro-parameters, the best-fit with the reverse-shock to the early afterglow 
has $\chi^2_\nu = 5.4$ for 136 dof, as it fails to account for the GeV prompt emission prior to 100 s, although
it explains well the early optical data and the X-ray data at 0.5--3 ks. We note that the reverse-shock magnetic 
parameter $\epsilon_B$ prior to 10 ks (from fitting the early optical afterglow) is 100 times larger than 
after 10 ks (from modelling for the radio emission). If the reverse-shock microphysical parameters were held 
constant across 10 ks, then the fit to the radio emission would have a $\chi^2_\nu$ twice larger, thus a decrease 
in $\epsilon_B$ at 10 ks is required. That may mean that the ejecta arriving at the blast-wave later are less magnetized.

\subsection{Other models with a more complicated afterglow medium structure}

 As discussed in \S\ref{sypeak}, to reconcile the radio and optical fluxes of afterglow 130427A requires that 
the peak of the forward-shock synchrotron crosses the optical after 3 ks. In turn, that requires (\S\ref{tenouswind})
a very weak stellar wind, about 300 less tenuous ($\dot{M}_{-5}/v_8 = 0.004$) than for the average Galactic WR star.
Then, the forward-shock radius is $R_a (3 \ks) = 1.5$ pc, while the wind bubble radius should be 
$R_s = 12\, (v_8^3/n_0)^{1/5}$ pc. Thus, if the circumstellar medium is sufficiently dense, it possible 
that $R_a (3 \ks) = R_s$. Alternatively, if the stellar wind had the average density, the wind termination 
shock could be encountered by the forward-shock at 3 ks, provided that the burst is embedded in a hot, 
highly pressurized environment (Chevalier et al 2004). At frequencies below the cooling break, the afterglow
light-curve should display a flattening when the forward-shock crosses the wind termination shock,
transiting from the $r^{-2}$ free wind to the quasi-homogeneous shocked wind.

 To be self-consistent, the interpretation of the 3 ks optical light-curve flattening as the blast-wave 
encountering the wind termination shock should attribute the entire afterglow emission to the same shock.
Then, the peak of the synchrotron spectrum must be below optical at all times when a decaying optical flux 
is measured, a model which overproduces radio emission, if the optical afterglow originates in the 
forward-shock (as shown in \S\ref{sypeak}). The subsequent steepening of the optical light-curve
at 20 ks cannot originate in the ambient medium stratification because, outside the termination shock, 
the shocked wind and circumstellar medium are still homogeneous. Instead, that light-curve steepening
should be attributed to the cooling frequency falling below the optical, which yields a steepening of 
the power-law flux decay by $\delta \alpha = 1/4$ (consistent with that measured for the optical light-curve 
of 130427A at 20 ks), and a softening of the optical spectrum by $\delta \beta = 1/2$ (consistent with 
the reddening reported by Perley et al 2013, after 10 ks).

 The fortuitous temporal coincidence of the cooling frequency falling below optical just after the
blast-wave arrives at the free-wind termination shock is not required if the discontinuity in the
ambient medium structure that yields the 3 ks optical light-curve flattening is caused by an internal
interaction within an unsteady stellar wind or by the interaction between the winds of two stars. 
In the former scenario, considered analytically by Chevalier \& Imamura (1983) and in the context of 
GRB afterglows by Ramirez-Ruiz et al (2005), a stronger wind produced by the WR star prior to its core-collapse
interacts with a slower wind ejected previously. In the latter scenario, proposed by Mimica \& Giannios (2011)
to be the source for more diverse afterglow light-curves, the GRB progenitor is in a dense stellar cluster,
where the mean distance between stars is below 1 pc, and the WR wind interacts with the weaker wind of 
a nearby O star or a later type. In either scenario, after the interaction with the shocked wind(s), 
which yields a light-curve flattening, the blast-wave goes into an $r^{-2}$ wind, which is the earlier 
WR wind or the wind of the nearby star, producing a light-curve steepening, with the flux decay index 
$\alpha$ returning to the value it had during the interaction with the free WR wind. Only the dense cluster 
scenario provides a natural explanation for the very weak wind inferred here from modelling the afterglow 
130427A: the wind of a B star located within 1 pc of the GRB progenitor. However, this scenario cannot 
explain why that weak wind extends over tens of pcs (as required by the duration of the afterglow, 
\S\ref{tenouswind}) despite the more powerful winds of nearby, earlier type stars.

 Thus, the 3 ks flattening and 20 ks steepening seen in the optical light-curve of 130427A could originate
from a forward-shock interacting with the more complex ambient medium resulting from an internal wind interaction
provided that microphysical parameters evolve such that this model does not exceed the 1--10 day radio 
measurements. Alternatively, radio emission is not overproduced if the entire afterglow emission arises
from the reverse-shock, and the optical light-curve flattening and steepening could result from the changing 
dynamics of the reverse-shock when the shocked-wind shell is crossed. Such light-curve features could also 
be due to variations in the density and Lorentz factor of the incoming ejecta, without any need for a 
non-uniform ambient medium. 

 However, a model where the entire afterglow emission arises from the same shock (reverse or forward) does 
not provide a natural explanation for the colour evolution displayed by 130427A, which becomes bluer after 
3 ks (Vestrand et al 2013), when the optical light-curve flattens, and redder after 10 ks (Perley et al 2013), 
when the optical light-curve steepens.
In contrast, the hybrid reverse-forward shock model explains naturally both the 3 ks spectral hardening, 
as due to the harder (below the synchrotron peak energy) forward-shock emission emerging from under the softer
reverse-shock emission, and the following spectral softening, caused by the peak energy of the forward-shock
synchrotron spectrum falling below optical.

\section{Conclusions}

 The closure relations expected between the forward-shock synchrotron flux decay index and spectral slope suggest 
a homogeneous ambient medium for GRB afterglow 130427A. Although long GRBs arise from massive stars that
drive powerful winds, a homogeneous medium is possible if the afterglow emission is produced in the shocked
wind. However, this afterglow's (10 ks) optical flux to (1--10 day) radio flux ratio and its slowly decaying 
radio light-curves disfavour that type of mbient medium. Instead, for an unevolving constant magnetic field parameter, 
the synchrotron spectrum peak flux is constant and the radio emission should have been brighter and slowly rising. 
With some fine-tuning of the evolutions of those micro-parameters, it may be possible to reduce the forward-shock 
model radio flux below measurements, while still accounting for the observed optical and X-ray light-curves.

 A wind-like medium ($n \propto r^{-2}$) is the more natural expectation for a massive star as the GRB progenitor.
The forward-shock emission still cannot account for the radio data because the expected radio light-curve is flat,
however, a wind-like medium yields a decreasing synchrotron spectrum peak flux, making it easier to keep the 
forward-shock radio emission below radio measurements. To explain the optical and X-ray flux decay after 
10 ks with the forward-shock synchrotron emission, a moderate energy injection into the forward-shock is required, 
increasing the shock energy by a factor 4 from 10 ks to 1 Ms. The agent of that energy injection should be some 
ejecta that arrive at the forward-shock at that time, which provides a natural explanation for the afterglow
radio emission: the reverse-shock that crosses the incoming ejecta.

 The reverse-shock must have been operating at even earlier times because the high early-optical to
late-radio flux ratio precludes a forward-shock origin of the optical afterglow emission prior to 10 ks.
Such a reverse-to-forward shock switch for the origin of the optical emission, occurring at few ks, is 
supported by the optical afterglow becoming bluer\footnotemark at that time (Vestrand et al 2013),
when the forward-shock emission, with a spectrum $F_\nu \propto \nu^{1/3}$ in the optical, begins to 
dominate the softer reverse-shock emission, with a spectrum $F_\nu \propto \nu^{-1/2}$.
\footnotetext{This feature, accompanied by a flattening of the optical flux decay, was previously observed in 
        two other GRB afterglows: 061126 (Perley et al 2008) and 080319B (Wo\'zniak et al 2009)}
As the peak of the forward-shock synchrotron spectrum falls below optical at about 10 ks, the optical
afterglow should become redder after 10 ks, as was observed by Perley et al (2013).

 However, for the reverse-shock to explain the 100 s -- few ks optical afterglow and the 1--10 day radio afterglow 
emission, the properties of the reverse-shock (microphysical parameters, kinetic energy and Lorentz factor of 
the incoming ejecta) must change around 10 ks. Furthermore, Vestrand et al (2013) have shown that the 
reverse-shock can also account for the optical flash (up to 100 s) and the GeV light-curve peak, but for 
microphysical different than after that peak.

 For this hybrid reverse-forward shock model, we find that the X-ray flux of GRB afterglow 130427A is accounted 
mostly by the forward-shock emission, from the tail of the first GRB pulse (50--100 s) up to 5 Ms, excluding 
the second GRB pulse at 100-500 s. The reverse-shock may have had a significant contribution 
to the early X-ray emission, at 500 s -- 2 ks. Both shocks give comparable GeV emissions. As shown in Fig
\ref{wind}, the radio emission from the forward-shock is expected to overshine that from the reverse-shock at 
30 day (or somewhat later, if energy injection continues after 1 Ms), yielding a flat flux $\siml 0.1$ mJy
until $\sim 200$ day, when the peak of the synchrotron spectrum falls below 10 GHz. If that flat radio flux 
is not seen, then the magnetic field parameter $\epsilon_B$ of the forward-shock must be decreasing, so that 
the peak flux of the forward-shock synchrotron spectrum falls-off faster than the $F_p \propto t^{-1/2}$ 
expected for $\epsilon_B = $const. 

 The relative dimness of the radio afterglow suggests that the peak of the synchrotron spectrum has crossed
the optical range at 10 ks. An immediate consequence is that the wind-like ambient medium is a factor 20 less
dense than the most tenuous wind measured for Galactic WR stars. 
We cannot provide a good argument for why GRB 130427A's progenitor had such a low mass-loss rate --to-- 
wind-speed ratio ($\dot{M}/v = 4\times 10^{-11} {\rm (M_\odot/yr)/(km/s)}$), but note that, owing to the weak 
wind, the afterglow remains highly relativistic and travels $\sim 100$ pc until the last observation epoch (50 day). 
For such a large afterglow radius to remain inside the free WR wind (i.e. within the wind termination shock), 
the GRB progenitor must have been embedded in a very tenuous medium, suggesting a supper-bubble blown by 
preceding supernovae and stellar winds.

 Owing to tenuous ambient medium, the afterglow transverse size, $2R_\perp = 2\Gamma ct \simeq 0.1 
(t/1 \day)^{3/4}$ pc, is unusually large, and implies a source apparent diameter of $\theta = 0.63\, 
(t/100 \day)^{3/4}$ mas, which may be resolved with radio interferometry.

 If the GeV emission of GRB afterglow 130427A arises from the forward-shock, then the up-scattering 
of the synchrotron emission occurred at the onset of the KN regime, where the reduction of 
the electron scattering cross-section lowers the Compton parameter, increases the synchrotron 
cooling-break frequency, and increases the synchrotron flux above that break (i.e. in the LAT range).
Furthermore, LAT must have measured the forward-shock inverse-Compton emission at photon energies 
above a few GeV. 

\vspace*{-5mm}
\section*{Acknowledgments}
 This work was supported by an award from the Laboratory Directed Research and Development programme
 at the Los Alamos National Laboratory and made use of data supplied by the UK Swift Science Data Centre 
 at the University of Leicester.

\end{document}